\documentclass[reprint, 
amsmath,amssymb, aps, prb]{revtex4-2}

\usepackage[colorlinks=true,linkcolor=blue]{hyperref}
\usepackage{natbib}
\usepackage{graphicx}
\usepackage{dcolumn}
\usepackage{bm}
\usepackage{amsmath}
\usepackage[T1]{fontenc}

\begin{document}
\preprint{APS/123-QED}
\title{The effect of iron layer thickness on the interlayer exchange coupling in Fe/MgO (001) superlattices}

\author{Anna L. Ravensburg}
\affiliation{Department of Physics and Astronomy, Uppsala University, Box 516, 75120 Uppsala, Sweden}

\author{Mat\'{i}as P. Grassi}
\affiliation{Department of Physics and Astronomy, Uppsala University, Box 516, 75120 Uppsala, Sweden}

\author{Bj\"orgvin Hj\"orvarsson}
\affiliation{Department of Physics and Astronomy, Uppsala University, Box 516, 75120 Uppsala, Sweden}

\author{Vassilios Kapaklis}
\affiliation{Department of Physics and Astronomy, Uppsala University, Box 516, 75120 Uppsala, Sweden}


\begin{abstract}

We describe the effect of the Fe layer thickness on the antiferromagnetic interlayer exchange coupling in [Fe/MgO]$_N$ superlattices. An increase in coupling strength with increasing Fe layer thickness is observed, which highlights the need of including the extension of both the layers when discussing the interlayer exchange coupling in superlattices.
\end{abstract}

\maketitle

\section{Introduction}

Epitaxial Fe/MgO heterostructures have sparked an intense research effort due to their large tunneling magnetoresistance \cite{Parkin2004, Yuasa2004}.
High crystal quality trilayer Fe/MgO/Fe heterostructures exhibit antiferromagnetic interlayer exchange coupling (IEC) \cite{FaureVincent2003,Ishigaki2010,Moubah2016,Magnus2018} mediated by spin-polarized tunneling through the MgO barrier \cite{FaureVincent2003, Slonczewski1989, Bruno1995}.
The coupling strength is exponentially decaying with increasing MgO layer thickness $t_\text{MgO}$ \cite{Katayama2006, Koziol2017, Bellouard2017}.
In [Fe/MgO]$_N$ superlattice structures with $N$ Fe/MgO bilayers, the interlayer exchange coupling has also been observed to exponentially decay with $t_\text{MgO}$ between 16.3 and 22.1~{\AA} MgO \cite{Moubah2016}.
Furthermore, in these epitaxial [Fe/MgO]$_N$ superlattices, discrete layer-by-layer magnetic switching is observed, resulting from the competition between the interlayer coupling and magnetocrystalline anisotropy \cite{Moubah2016}.
However, understanding and tuning the sequential magnetic switching in these heterostructures and the precise mechanism governing it has been proven challenging \cite{Magnus2018, Warnatz2021}.
The MgO layer thickness, impurities or defects in the MgO layer, temperature, and growth conditions are important parameters affecting the interlayer coupling \cite{FaureVincent2003, Ye2005, Katayama2006, Bellouard2008, Chiang2009, Moubah2016, Bellouard2017, Magnus2018}.
Oxygen vacancies \cite{Ye2005, Katayama2006}, along with magnetic impurities which might be present in the MgO layers \cite{Chiang2009} have also been shown to be relevant.
Finally, a non-trivial dependence of the coupling strength in [Fe/MgO]$_N$ superlattices on the number of bilayer repetitions, i.e., the extension of the whole stack has been reported \cite{Warnatz2021}.

Systematic variations of the MgO layer thickness in previous works \cite{Katayama2006, Koziol2017, Bellouard2017, Moubah2016} relate to the modification of the tunneling barrier.
On the other hand, the study of the dependence on $N$ in superlattices, highlights potential collective effects between quantum well states in the Fe layers making up the [Fe/MgO]$_N$ superlattices \cite{Warnatz2021}.
A detailed investigation of the Fe layer thickness dependence, would thus shed light on the importance of the extension of the quantum wells formed in the Fe layers for the interlayer coupling strength.
To this end, we investigate the influence of the Fe layer thickness on the IEC and sequential switching in [Fe/MgO]$_N$ superlattices.
The Fe layer thickness was varied while the MgO layer thickness was kept constant.
We present a detailed characterization of the layering and crystal structure of the superlattices, employing x-ray scattering techniques, and report on the increase of the IEC with increasing Fe layer thickness.

\section{Methods}

Superlattices of [Fe/MgO]$_N$ with $N=$ 8 and 10 bilayer repetitions were deposited on single crystalline MgO~($001$) substrates by direct current (dc) and radio frequency (rf) magnetron sputtering.
Within one deposition process, identical copies were grown on a 10$\times$10~mm$^2$ and a 20$\times$20~mm$^2$ sized substrate, eliminating this way all uncertainties concerning differences in growth conditions.
The thicknesses of the MgO layers were kept constant at $t_\text{MgO}$ = 17(1) {\AA}, while the thickness of the Fe layers was chosen to vary in the range of $t_\text{Fe}$ = 11 to 23~{\AA}.
Prior to the deposition, the substrates were annealed in vacuum at 1273(2)~K for 600~s.
The base pressure of the growth chamber was below 5$\times$10$^{-7}$~Pa.
The target-to-substrate distance in the deposition chamber was approximately 20~cm.
The depositions were carried out in an Ar atmosphere (gas purity $\geq$~99.999~\%, and a secondary getter based purification) from an elemental Fe (50~W, dc, diameter: 5.08~cm) and an MgO compound target (50~W, rf, diameter: 5.08~cm) at 438(2)~K and 0.67~Pa.
Target power and deposition temperature were optimized previously with respect to well-defined layering in combination with the highest crystal quality.
In order to prevent surface oxidation of the films, the samples were capped at ambient temperature ($<$~313(2)~K) and 1.07~Pa with 50~{\AA} Pt (50~W, dc, diameter: 5.08~cm).
The targets were cleaned by sputtering against closed shutters for at least 60~s prior to the deposition of each layer. 
The deposition rates (Fe: 0.1~{\AA}/s, MgO: 0.01~{\AA}/s, Pt: 0.8~{\AA}/s) were calibrated prior to the growth of the samples.
In order to ensure thickness uniformity, the substrate holder was rotated at 30~rpm during the deposition.

\begin{figure*}[ht!] 
\centering
\includegraphics{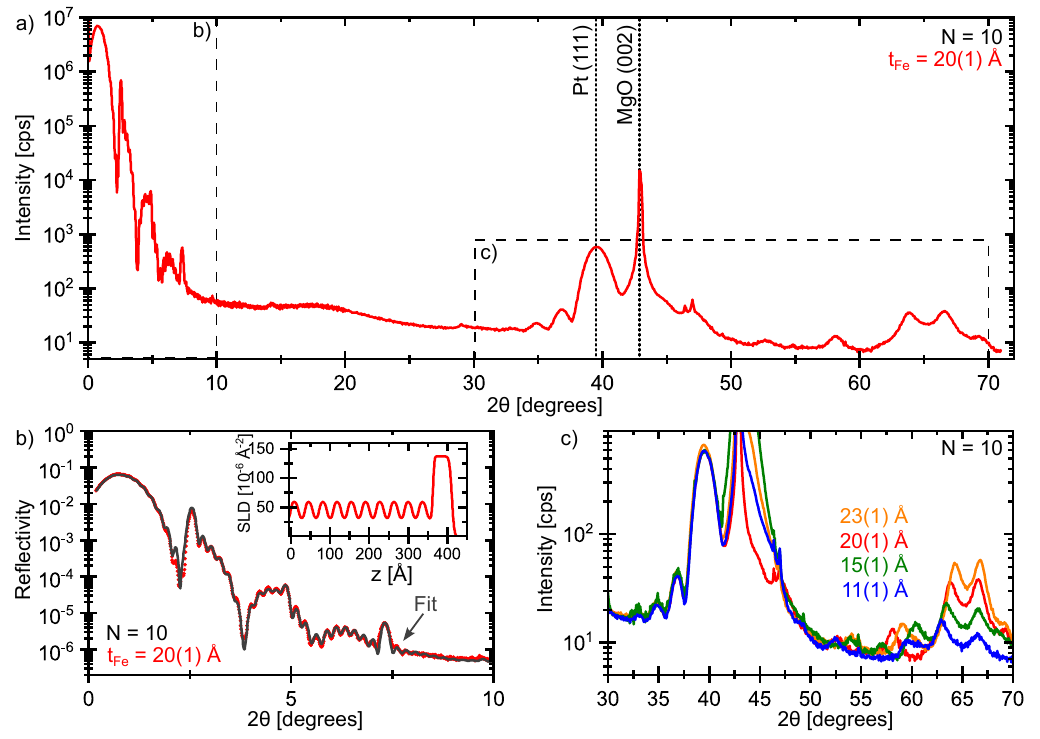}
\caption{a) X-ray scattering pattern of a [Fe/MgO]$_{10}$ superlattice with 20(1)~{\AA} Fe layer thickness. In b) x-ray reflectivity is displayed separately including a fit. The scattering length density profile based on the fit is displayed in the inset. c) X-ray diffraction of [Fe/MgO]$_{10}$ superlattice samples with different Fe layer thicknesses.} 
\label{fig:xrays}\hfill
\end{figure*}

X-ray reflectometry (XRR) and diffraction (XRD) were carried out using a Bede D1 diffractometer equipped with a Cu K$_{\alpha_1}$ x-ray source operated at 35~mA and 50~kV.
For monochromatizing the beam by reducing the Cu K$_\beta$ radiation, the setup included a G\"obel mirror and a Ni filter on the incidence and detector side, respectively.
A circular beam mask (diameter: 0.005~m), incidence and detector slits (both 0.0005~m) were also used.
The x-rays were detected with a Bede EDRc x-ray detector.
The instrument angles for the coupled $2\theta$-$\theta$ scans were aligned to the sample surface for XRR and to the [Fe/MgO] crystal planes for XRD measurements.
The measured XRR data was fitted using \textsc{GenX} \cite{Bjorck2007, Glavic_Bjorck_2022} enabling the determination of the scattering length density (SLD) profile, which includes information on layer thickness and roughness.
Rocking curve measurements in XRD were fitted with a Pseudo-Voigt function to mimic the effect of mosaic spread.

Magnetization measurements were performed at ambient temperature using a longitudinal magneto-optical Kerr effect (L-MOKE) setup with $s$-polarized light.
The magnetic response was measured parallel to an in-plane applied magnetic field along the Fe magnetic easy and hard axes, i.e. Fe~[$100$] and Fe~[$110$], respectively.
The data was averaged over 10 full loop recordings.
Additional measurements were conducted in an Evico Magnetics KerrLab Kerr microscope in longitudinal and transverse mode.
The determined saturation fields are corrected for the coercivity.
The values of the saturation field along magnetic easy and hard axes were determined by the intersection of the hysteresis loop with two straight lines at $M=M_S$ (the hard axis case is illustrated in the SM).
Displayed error bars of the saturation fields and the remanence magnetization correspond to the standard deviation of the extracted parameters from the two symmetrical axes (90-degrees sample rotation), two field-scan directions, and two field polarities. 

\section{Results and Discussion}
\subsection{Layering and crystal structure}

\begin{figure*}[ht!] 
\centering
\includegraphics{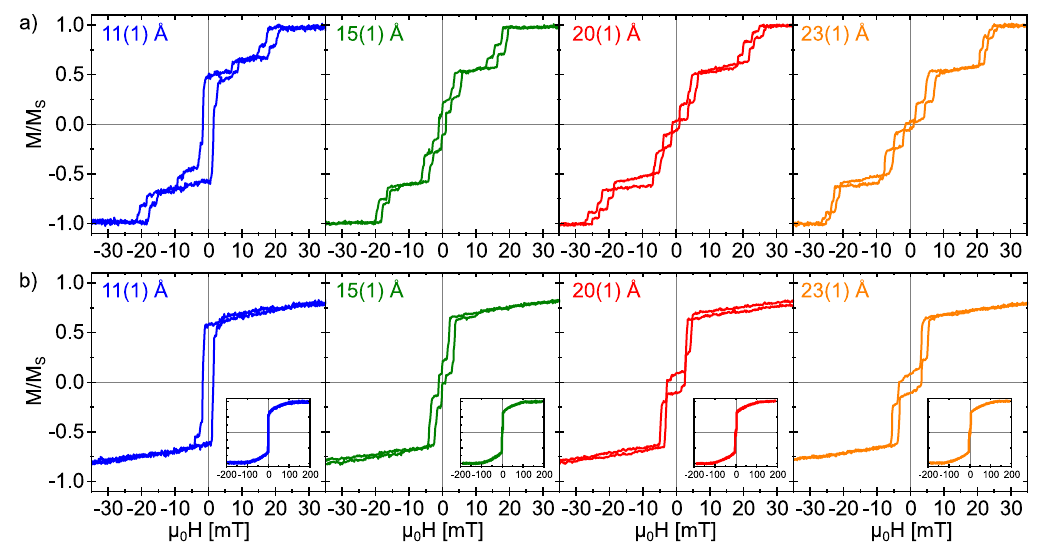}
\caption{Magnetic hysteresis loops of [Fe/MgO]$_{10}$ superlattices measured with L-MOKE on 10$\times$10~mm$^2$ samples with varying Fe layer thickness with an applied field along the a) Fe magnetic easy axis, i.e., Fe~[$100$] and b) Fe magnetic hard axis, i.e., Fe~[$110$].} 
\label{fig:loops}\hfill
\end{figure*}

A full x-ray scattering pattern of an [Fe/MgO]$_{10}$ superlattice with 20(1)~{\AA} Fe and 17(1)~{\AA} MgO layer thicknesses is displayed in Fig.~\ref{fig:xrays}a.
Below 10~degrees in 2$\theta$, pronounced Kiessig fringes \cite{Kiessig1931} are observed, consistent with the presence of flat and distinct layering with low layer roughness (below 5(1)~{\AA}), which is confirmed in the scattering length density (SLD) profile based on a fit displayed in Fig.~\ref{fig:xrays}b.
All nominal Fe and MgO layer thicknesses within this study were determined using XRR.
Analysis of the results confirmed that the MgO layer thickness was the same, 17(1)~{\AA}, while the thicknesses of the Fe layers were determined to be 11(1)~{\AA}, 15(1)~{\AA}, 20(1)~{\AA}, and 23(1)~{\AA}. 
Henceforth, all presented thicknesses correspond to the experimentally determined thicknesses of the Fe and MgO layers.

Diffraction peaks at around 39 and 43~degrees in 2$\theta$ are observed, arising from the Pt capping and the MgO~($001$) substrate, respectively. 
Laue oscillations around the Pt Bragg peak are obtained for all the samples, see Fig.~\ref{fig:xrays}c, providing conclusive evidence for well-defined texture and thickness of the Pt layer \cite{Ravensburg2023Fit}.
Fe/MgO superlattice satellite peaks are visible between 52 and 70~degrees in 2$\theta$.
Their spacing and positions relate to the Fe/MgO bilayer thickness and, hence, vary with Fe layer thickness, which was confirmed with x-ray scattering intensity simulations employing GenL \cite{Ravensburg2023Fit}.
Their intensity relates to the amount of scattering material, i.e., the total Fe and MgO layer thicknesses.
Furthermore, the satellite peak intensity is elevated by the broad Fe~($002$) Bragg peak, which is observed at around 65 degrees in 2$\theta$.
The Bragg peak intensity relates to the total Fe layer thickness and is, therefore, higher for the samples with thicker Fe layers.
No Laue oscillations are visible between the superlattice satellite peaks for [Fe/MgO]$_{8}$ and [Fe/MgO]$_{10}$.
This is attributed to the mismatch between the layers, giving rise to defects causing incoherent scattering from the Fe atomic planes \cite{Muehge1994, Ravensburg2022}.
The out-of-plane atomic distances in Fe and MgO are 2.866~{\AA} \cite{Vassent1996} and 4.212~{\AA} \cite{LandoltBornstein1994}, respectively.
Hence, the atomic step heights in the Fe and MgO lattices are incommensurate, yielding finite terrace widths in the superlattice, which can suppress both, diffraction peaks as well as Laue oscillations \cite{Raanaei2008, Ravensburg2023Fit}.
This is in stark contrast to, e.g., the growth of [Fe/V]$_N$~(001) superlattices, within which the layers have similar lattice parameters \cite{ISBERG1997483}.
Furthermore, the mismatch results in biaxial elastic in-plane strain in both Fe and MgO, which may cause a tetragonal distortion of the cubic unit cells.
Strain in Fe has an impact on its magnetic properties \cite{Friak2001}.
Moreover, dislocations for relaxation at the critical thickness, which is around 20~{\AA} for Fe~($001$) on MgO~($001$) \cite{Ravensburg2022}, can affect the magnetic properties.
Defects may also originate from substrate twinning, which is common in commercially available MgO substrates \cite{Schroeder2015}.
More twins and larger mosaic spreads were observed for 20$\times$20~mm$^2$ compared to 10$\times$10~mm$^2$ substrates.
In specular scans for samples on 20$\times$20~mm$^2$ substrates, multiple reflections attributed to asymmetric substrate peaks were observed, indicating defects in the single crystalline structure.
This difference in crystal quality between the substrates is found to alter the magnetic properties in [Fe/MgO]$_{N}$ superlattices (see Supplemental Material (SM)).
Therefore, a comparison between samples, in particular for different (substrate) batches, has to be done with care, as observations in magnetic properties need to be related to the crystal structure.

\subsection{Magnetic characterization}


Magnetic hysteresis loops of [Fe/MgO]$_{10}$ superlattices, measured with an applied magnetic field along the Fe easy axis, are displayed in Fig.~\ref{fig:loops}a.
Kerr microscopy revealed domain wall nucleation and motion during the magnetization reversal, as expected in sequential switching of the magnetic layers.
The magnetic domain size in the Fe layers was determined to lie in the order of several hundreds of micrometers to millimeters \cite{WarnatzThesis} and is, thus, comparable to the probing area of L-MOKE measurements.
No dependence of the domain size on the Fe layer thickness was found. 
Hence, sequential switching of the layers in the superlattices is apparent from the presence of discrete steps in the field response, which is assumed to stem from a single or few domains in each Fe layer.
The findings are in line with previous reports \cite{Moubah2016, Magnus2018}.
Coming from positive saturation, the first switching occurs at a positive applied field, in agreement with an antiferromagnetic IEC of the Fe layers in all samples \cite{Moubah2016, Magnus2018}.
The hysteresis loops obtained for the samples having 8 bilayers, are shown in the SM.

\begin{figure}[ht!] 
\centering
\includegraphics{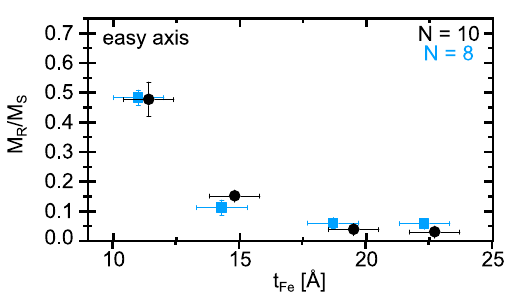}
\caption{Remanent magnetization $M_R$ normalized to the saturation magnetization $M_S$ displayed over Fe layer thickness $t_\text{Fe}$ for [Fe/MgO]$_N$ superlattices with 8 and 10 bilayer repetitions $N$ on 10$\times$10~mm$^2$ substrates for an applied magnetic field along the Fe easy axis.} 
\label{fig:rem}\hfill
\end{figure}
The remanent configuration changes with Fe layer thickness.
For the sample with the thinnest Fe layers, 11(1)~{\AA}, the remanent state corresponds to 0.48(6)$\times$M$_S$, consistent with every other layer pointing along the sensitivity axis and the others oriented perpendicular to that \cite{Magnus2018, Warnatz2021}.
The inferred 90~degrees configuration could either be enabled by the presence of an activation barrier for the switching or by the presence of a biquadratic coupling \cite{Chiang2009}.
Biquadratic coupling is commonly associated with coupling in systems with metal spacer layers, which varies between antiferromagnetic and ferromagnetic with varying spacer layer thickness.
Biquadratic coupling may be attributed to an extrinsic fluctuation mechanism leading to frustration of the bilinear exchange coupling \cite{Slonczewski1993, Demokritov1998}.
However, for insulating spacers, the coupling is reported to be only antiferromagnetic \cite{Bruno1995}.
A 90~degrees remanent configuration has been observed for [Fe/MgO]$_{10}$ superlattices with thicker MgO barriers of 19.6~{\AA}, which was attributed to the coupling strength being smaller than the nucleation field of the domain reversal in the samples \cite{Magnus2018}.
In contrast, the remanent magnetization in the samples with larger Fe layer thickness is closer to being compensated, i.e., 0.151(3)$\times$M$_S$, 0.039(4)$\times$M$_S$, and 0.031(5)$\times$M$_S$ for the samples with 15(1), 20(1), and 23(1)~{\AA} Fe layer thickness, respectively.
A compensated remanent state corresponds to a fully antiferromagnetic alignment and was reported for [Fe/MgO]$_{10}$ with thinner MgO barriers of 16.4~{\AA} \cite{Magnus2018}, having larger coupling strength as compared to the thicker layers \cite{Moubah2016}.
In these samples, the coupling between the Fe layers seems however sufficiently strong to drive the order to an antiferromagnetic remanent state.
The normalized remanent magnetization $M_R/M_S$ is plotted as a function of Fe layer thickness in Fig.~\ref{fig:rem} for [Fe/MgO]$_{8}$ and [Fe/MgO]$_{10}$.
Independent of whether a superlattice consists of 8 or 10 bilayer repetitions, the remanent magnetization decreases with increasing Fe layer thickness.

Besides the coupling strength, different strain states can affect the magnetic properties of Fe \cite{Friak2001} and defects can act as pinning points for domain wall motion.
In Fe layers exceeding the critical thickness of around 20~{\AA}, a higher defect density is expected.
However, no signs of domain wall pinning in samples with thicker Fe layers is observed.
The metastable 90~degrees configuration is observed for samples with thinner Fe layers and based on the hysteresis loops displayed in Fig.~\ref{fig:loops}a, the coercive field is not found to be increasing with increasing Fe layer thickness.
Hence, the influence of different strain states on the remanent configuration is neglected in the further analysis.

\begin{figure}[t!] 
\centering
\includegraphics{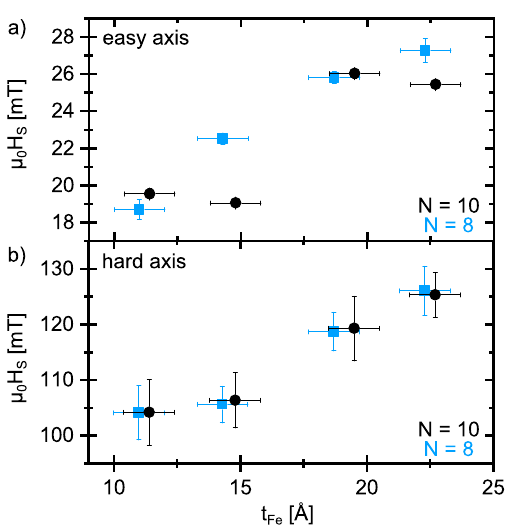}
\caption{Saturation field $H_S$ displayed over Fe layer thickness $t_\text{Fe}$ for [Fe/MgO]$_N$ superlattices with 8 and 10 bilayer repetitions $N$ on 10$\times$10~mm$^2$ substrates for an applied magnetic field along a) the Fe easy axis, b) the Fe hard axis.} 
\label{fig:sat}\hfill
\end{figure}

The same relative change in remanent magnetization $M_R/M_S$ with Fe layer thickness is observed in loops measured with an applied field along the Fe magnetic hard axis, which are displayed in Fig.~\ref{fig:loops}b for [Fe/MgO]$_{10}$ superlattices.
Here we notice that more than four times larger fields are required to overcome the crystalline anisotropy as compared to the saturation field along the easy axis of the samples.
Differences in the switching are also noticed for the two magnetic field directions, where discrete steps are observed up until 2 to 5~mT, when the field is applied along the hard axis of the samples.
The magnetization at these fields is approximately 0.63$\times$M$_S$, indicating a mixture of switching and coherent rotation of layers.
Hence, when the applied field is along the easy axis, the individual layers do switch, one or more at the time, which is not observed when the field is applied along the hard axis.
Above the threshold, the change in the alignment of magnetization of the Fe layers appears to be dominated by coherent rotation for fields along the hard axis. 
The saturation fields along the easy and hard axes, $H_S^{\text{EA}}$ and $H_S^{\text{HA}}$, were extracted from the hysteresis loops.
The results are shown in Fig.~\ref{fig:sat}.
As seen in the figure, the saturation field increases with increasing Fe layer thickness in both cases. 

%
%
%
%
%
%


To obtain quantitative determination of $J$ and $K$, we define the energy density $E_i$ of the layer $i$ as:
\begin{equation}
\label{equ:E}
    \begin{split}
    E_i =& - \mu_0 M_S H\,  \mathrm{cos}(\theta_i) + \frac{K}{4}\, \mathrm{sin}^2(2\theta_i) \\
         &- J\, \mathrm{cos}(\theta_i-\theta_{i+1})- J\, \mathrm{cos}(\theta_i-\theta_{i-1}),
    \end{split}
\end{equation}
where $M_S=1.71$~kA/m is the saturation magnetization \cite{Crangle1971}, $\mu_0$ is the vacuum permeability, $H$ is the applied field, and $\theta_i$ is the angle of the magnetization of the layer $i$ measured from the easy axis.
For the outermost layers, with only one neighboring Fe layer present, one of the last two terms vanishes.
With this definition, $J$ is negative for an antiferromagnetic IEC.
\begin{figure}[t!] 
\centering
\includegraphics{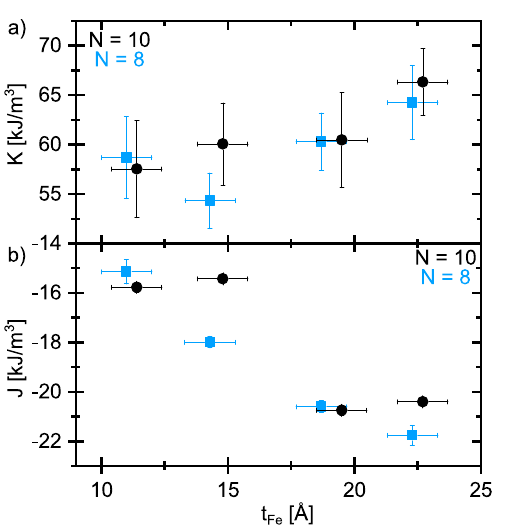}
\caption{a) Fe cubic anisotropy $K$ and b) volumetric coupling strength $J$ displayed over Fe layer thickness $t_\text{Fe}$ for [Fe/MgO]$_N$ superlattices with 8 and 10 bilayer repetitions $N$ on 10$\times$10~mm$^2$ substrates determined from hysteresis loops measured with an applied magnetic field along the Fe easy and hard axes.} 
\label{fig:J_K}\hfill
\end{figure}
From this expression, it is possible to calculate the saturation field along the hard axis $\ H_S^{\text{HA}}$ (see SM) as:
\begin{equation}
\label{equ:H_HA}
\,H_S^{\text{HA}} = \frac{2\,K-4\,J}{\mu_0 \,
M_S}.
\end{equation}

The dependence of the saturation field along the easy axis $H_S^{\text{EA}}$ as a function of $K$ and $J$ is not accessible in an analytical form for a multilayer structure \cite{Bloemen1994,Dieny1990} and has therefore to be extracted from simulations of the data.
The functional dependence of $H_S^{\text{EA}}(K, J)$ was calculated, and it was found that it decays with $K$, converging to $H_S^{\text{EA}}=-2J/M_S$ in the limit of large $K/|J|$ (for details see SM).
With both experimental values for $ H_S^{\text{HA}}$, $H_S^{\text{EA}}$ and Eq.~\ref{equ:H_HA} as well as $H_S^{\text{EA}}(K, J)$ (see SM), it is possible to calculate the values of the coupling strength $J$ and the Fe cubic anisotropy $K$, which are shown in Fig.~\ref{fig:J_K}.
The anisotropy is determined to be weakly changing with Fe thickness (60(4)~kJ/m$^3$) in the range studied here. 
For 20~{\AA} Fe on MgO, hence in a comparable thickness range, a $K=56$~kJ/m$^3$ was reported \cite{Postava1997}.
Moreover, in line with the findings about the remanent state being affected by the Fe layer thickness, the antiferromagnetic coupling strength is found to increase with Fe layer thickness, independent of the number of bilayer repetitions.
\vspace{5mm}

\section{Summary}

The antiferromagnetic coupling strength [Fe/MgO]$_N$ superlattices with 17~~{\AA} thick MgO layers is found to linearly increase with Fe layer thickness, in the range between 11 and 23~{\AA}.
The Fe cubic anisotropy was determined to be weakly increasing in the same thickness range, being close to the value reported for Fe single layers.
These findings indicate that the coupling in [Fe/MgO]$_N$ superlattices is not solely defined by the thickness of the MgO barriers, but along with the number of bilayer repetitions $N$ \cite{Warnatz2021}, the thickness of the individual Fe layers is also an essential parameter.
This calls for a closer look on the coupling mechanism and effects in such superlattices, highlighting the importance of all length scales present in these ($t_\text{MgO}$, $t_\text{Fe}$, and $N$($t_\text{MgO}+t_\text{Fe}$)).
These observations challenge our current understanding of electronic and spin effects in magnetic quantum well structures comprised of metal-oxide heterostructures, with potential impact on emergent technologies employing their magnetic properties. 

\begin{acknowledgments}
VK would like to acknowledge financial support from the Swedish Research Council (Project No. 2019-03581). MPG and VK also acknowledge support from the Carl Trygger Foundation (Project No. CTS21:1219).
\end{acknowledgments}
\vspace{5mm}

\section*{Data availability}
The data that support the findings of this study are available from the authors upon reasonable request.

\providecommand{\noopsort}[1]{}\providecommand{\singleletter}[1]{#1}%
%

\pagebreak
\onecolumngrid
\newpage
\begin{center}
\textbf{\large Supplemental Material: The effect of iron layer thickness on the interlayer exchange coupling in Fe/MgO (001) superlattices}
\end{center}
\setcounter{equation}{0}
\setcounter{figure}{0}
\setcounter{table}{0}
\setcounter{page}{1}
\setcounter{section}{0}
\makeatletter
\renewcommand{\theequation}{S\arabic{equation}}
\renewcommand{\thesection}{S-\Roman{section}}
\renewcommand{\figurename}{Supplementary FIG.}
\renewcommand{\thefigure}{{\bf \arabic{figure}}}
\renewcommand{\bibnumfmt}[1]{[S#1]}
\renewcommand{\citenumfont}[1]{S#1}
\renewcommand{\thepage}{S-\arabic{page}}

\section{Magnetic hysteresis loops of superlattices with 8 bilayer repetitions}

The magnetic hysteresis loops of [Fe/MgO]$_{8}$ superlattices with varying Fe layer thickness, measured with an applied field along the Fe magnetic easy and hard axis, are shown in Supplementary Fig.~\ref{fig:loops8}a and b, respectively.
Qualitatively, the loops exhibit the same shape as the loops recorded for superlattices with 10 bilayer repetitions.
All extracted values for relative remanent magnetization $M_R/M_S$ and saturation fields $\mu_0 H_S^{EA}$ and $\mu_0 H_S^{HA}$ are displayed in the main text.
\begin{figure*}[ht!] 
\centering
\includegraphics[width=\textwidth]{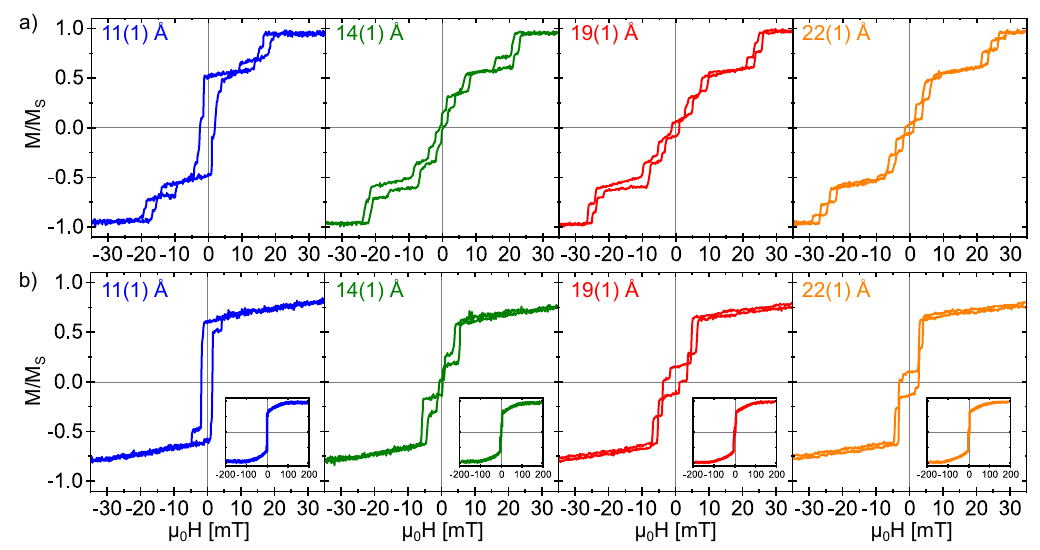}
\caption{Magnetic hysteresis loops of [Fe/MgO]$_{8}$ superlattices measured with L-MOKE on 10$\times$10~mm$^2$ samples with varying Fe layer thickness with an applied field along the a) Fe magnetic easy axis, i.e., Fe~[$100$] and b) Fe magnetic hard axis, i.e., Fe~[$110$].} 
\label{fig:loops8}\hfill
\end{figure*}
\vspace{2cm}

\section{Dependence of the magnetic properties on the substrate size}

All analysis presented in the main text was conducted on [Fe/MgO]$_{N}$ superlattices grown on 10$\times$10~mm$^2$ MgO~($001$) substrates.
However, copies of the deposited [Fe/MgO]$_{10}$ superlattices on 20$\times$20~mm$^2$ MgO~($001$) substrates show differences in the magnetization reversals, displayed in Supplementary Fig.~\ref{fig:loops20}, despite an identical layering and a similar crystal structure.
All 20$\times$20~mm$^2$ samples exhibit a remanent state around 0.5(0)$\times$M$_S$ corresponding to a 90~degrees orientation of the magnetization in the Fe layers at remanence.
The sample with 15(1)~{\AA} Fe layer thickness shows a slightly lower remanent magnetization but exhibits 0.5(0)$\times$M$_S$ already at applied fields below 1~mT.
A strong dependence of the shape of the magnetic hysteresis loop on the measured position on the sample was found.
Furthermore, the saturation field of the [Fe/MgO]$_{10}$ superlattices on 20$\times$20~mm$^2$ MgO~($001$) substrates is significantly lower and in the order of 10 to 15(1)~mT, compared to their copies on 10$\times$10~mm$^2$ substrates.
Based on the previous analysis of remanent state and saturation field, the coupling strength in the samples on 20$\times$20~mm$^2$ is smaller compared to their 10$\times$10~mm$^2$ counterparts.
Nevertheless, the trends observed for the saturation fields with Fe layer thickness were found to be valid for both, 20$\times$20~mm$^2$ and 10$\times$10~mm$^2$ samples in relative comparison.

\begin{figure*}[t!] 
\centering
\includegraphics[width=\textwidth]{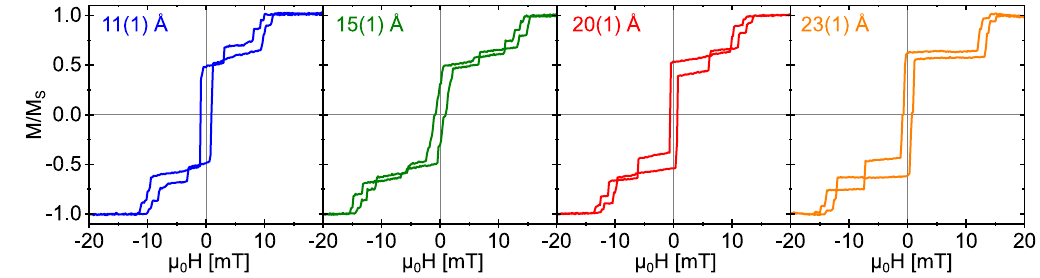}
\caption{Magnetic hysteresis loops of [Fe/MgO]$_{10}$ superlattices measured with a Kerr microscope on 20$\times$20~mm$^2$ samples with varying Fe layer thickness with an applied field along the Fe magnetic easy axis, i.e., Fe~[$100$].} 
\label{fig:loops20}\hfill
\end{figure*}

The change in magnetic properties is attributed to differences in the substrate quality.
As the switching mechanism in [Fe/MgO]$_{N}$ superlattices is found to rely on domain wall movements, crystallographic defects may inhibit this motion and lower the coupling strength between the layers.
More surface area of an MgO~($001$) substrate increases the probability for the presence of crystallographic twins and an increase in mosaic spread, i.e., crystal misorientation, in line with the analysis of the crystal structure of the superlattices on these substrates, presented above.
Therefore, the coupling strength in [Fe/MgO]$_{N}$ superlattices is found to be highly dependent on substrate quality and crystal structure/quality.

\section{Dependence of the easy-axis saturation field on $K$ and $J$}

The dependence of $H_S^{EA}$ as a function of the antiferromagnetic IEC $J$ and the cubic anisotropy $K$ only becomes trivial in the limit of large $K$ \cite{Dieny1990,Folkerts1991}, where: 
\begin{equation}
    H_S^{\text{EA}}=\frac{2\,J}{\mu_0\,M_S}.
    \label{eq:J_2}
\end{equation}

In this limit, the switching process is modelled by a first-order transition where the magnetization of one layer is exactly 90 or 180 degrees from the direction defined by the applied field.
This process is known as spin-flip.
When the cubic anisotropy is weaker, the magnetization of the different layers can explore a broader range of configurations in order to minimize the energy.

In the case where the magnetization of one layer rotates to $\theta=\pi/2$, the system can further minimize its energy if the magnetization of the two neighboring layers drives away from saturation.
This could be understood from the large slope of the term $J\,\mathrm{cos}(\theta_i - \theta_{\pm1})$ in Eq.~1 in the main text, when $\theta_i - \theta_{\pm1}=\pi/2$, and the small slope of $\mu_0 H M_S\, \mathrm{cos}(\theta_{\pm1})$ when $\theta_i=0$. 
Therefore, it is expected that the value of $H_S^{\text{EA}}$ depends on both $K$ and $J$ as $H_S^{EA}=F(J,K)$.
This dependence can be written as $h=\frac{f(k)}{\mu_0 M_S}$, where $h=H_S^{\text{EA}}/|J|$ and $k=K/|J|$ are adimensional parameters.
In order to explore this dependence, Eq.~1 in the main text was used to determine the angle $\theta_{\pm1}$, which minimizes the energy, and the field at which the spin-flip becomes energetically favorable.
The minimal model to study the deviation from Eq.~ \ref{eq:J_2} requires considering 5 layers.
To calculate the reduced saturation field $h$, a switched state is proposed, and its energy is compared with the saturated state.
In the switched configuration, the angles $\theta_i$ of the magnetization of the layer $i$ are:
\begin{equation}
\begin{split}
    \theta_1=\theta_5=0,\\
    \theta_2=\theta_4=\delta,\\
    \theta_3=\pi/2.    
\end{split}
\end{equation}
It must be noted that if $\theta_3$ is unconstrained, it always minimizes the energy at the exact angle of $\pi/2$.
By taking a step in $\delta$ of 0.009 rad, and 0.0011 in $h$ it was possible to obtain $f(k)$. 
A similar procedure was followed to calculate the reduced field related to a spin-flop transition (second-order transition), where the system transitions from saturation to a configuration with $\theta_{\text{odd}}=-\theta_{\text{even}}$.
In this case, only two layers have to be considered.
However, to take the boundary conditions imposed by the outermost layers into account, the IEC term in Eq.~1 in the main text has to be multiplied by a factor $(N-1)/N$, where $N$ is the amount of layers of the superlattice.
The results for both spin-flip and spin-flop are shown in Supplementary Fig.~\ref{fig:f(k)}.
For values under $k=1$, spin-flop dominates at higher fields.

\begin{figure*}[h!] 
\centering
\includegraphics{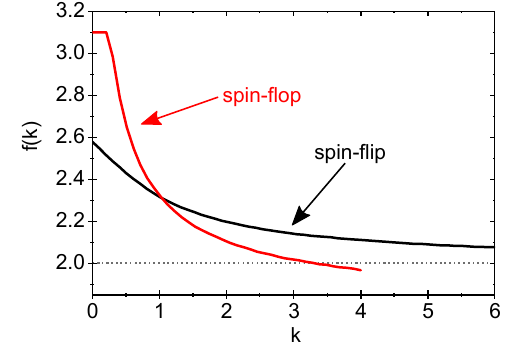}
\caption{Calculated values of $f$ as a function of the reduce parameter $k=K/|J|$. For each $k$, the highest value dominates the transition from saturation.} 
\label{fig:f(k)}\hfill
\end{figure*}

It can be observed that in the limit of large $k$ the factor $f(k)$ for spin-flip converges to 2 in agreement with Eq.~\ref{eq:J_2}.
From \citet{Postava1997} and Eq.~\ref{eq:J_2} is it possible to estimate that $k$ should lie close to 3.
Therefore, $f(k)$ can be approximated in the range $k\in [2.8;\,3.9]$ by a linear function $f_l(k)=A\,k+B$.
With this simplified expression, the saturation field can be calculated as:
\begin{equation}
    H_S^{\text{EA}}=\frac{A\,K+B\,J}{\mu_0\,M_S},
    \label{eq:saturation_EA}
\end{equation}
where $A=-$0.04 and $B=-$2.26.

\section{Dependence of the hard-axis saturation field on $K$ and $J$}

For extracting values for the coupling strength from hard axis hysteresis loops, a ferromagnetic multilayer with a fourfold magnetocrystalline anisotropy is considered.
The interlayer exchange coupling is antiferromagnetic ($J<0$). 
When the external magnetic field $\bm{H}$ is reduced from saturation, each layer system goes through a second-order transition at the critical field $H_S^{HA}$, as shown in Supplementary Fig.~\ref{fig:loops8}b or in the main text.
At lower fields, the angle $\theta_i$ between the magnetization of each layer $\bm{M}_i$ and the magnetic field takes a non-zero value.
Therefore, the total magnetization along the field direction can be calculated as $\frac{M_S}{N}\sum^N_i \mathrm{cos}(\theta_i)$, where $N$ is the number of layers.

However, if $N>2$, the layers with two neighbors present a higher saturation field $ H_S^{HA}$ which can be extracted from the measured hysteresis loops.
For each of these "middle" layers, if even and odd layers angles satisfy $\theta=-\theta$, the energy density per layer is reduced to: 
\begin{equation}
    E = -\mu_0 M_S H\,\mathrm{cos}(\theta) - \frac{K}{4}\,\mathrm{sin}^2(2\theta) -J\,\mathrm{cos}(2\,\theta).
    \label{E_minima}
\end{equation}
From here, the minima can be found from:
\begin{equation}
    \frac{\partial E}{\partial \theta} = \mu_0 M_S H\,\mathrm{sin}(\theta) -K\,\mathrm{sin}(2\theta)\,\mathrm{cos}(2\theta)+ 2J\,\mathrm{sin}(2\,\theta)=0.
\end{equation}
Using $\mathrm{sin}(2\theta)=2\,\mathrm{sin}(\theta)\mathrm{cos}(\theta)$, this equation can be written as:
\begin{equation}
    \mu_0 M_S H\,\mathrm{sin}(\theta) - 4K\,\mathrm{sin}(\theta)\mathrm{cos}(\theta)\,\mathrm{cos}(2\theta) + 4J\,\mathrm{sin}(\theta)\mathrm{cos}(\theta)=0.
\end{equation}
One trivial solution is $\theta=0$ and it represents the ground state at $H> H_S^{HA}$.
We know that below the saturation field, this minimum will no longer be the ground state and the system will transition to a solution with $\theta \neq 0$.
However, at the critical field, this second solution should present $\theta=0$.
Therefore, by discarding the trivial solution and imposing $\theta=0$ we obtain $\mu_0 M_S H - 8K + 4J =0$, and the critical field is calculated as:
\begin{equation}
    H_S^{\text{HA}} = \frac{ 2\,K - 4\,J}{\mu_0 M_S}.
    \label{eq:saturation_HA}
\end{equation}
At this field, all the layers are expected to go through a soft transition.
From this expression and Eq.~\ref{eq:saturation_EA}, it is possible to calculate the values of $J$ and $K$ from $H_S^{\text{HA}}$ and $ H_S^{\text{EA}}$.

\section{Experimental determination of $H_S^{\text{HA}}$}

\begin{figure*}[h!] 
\centering
\includegraphics[width=0.8\textwidth]{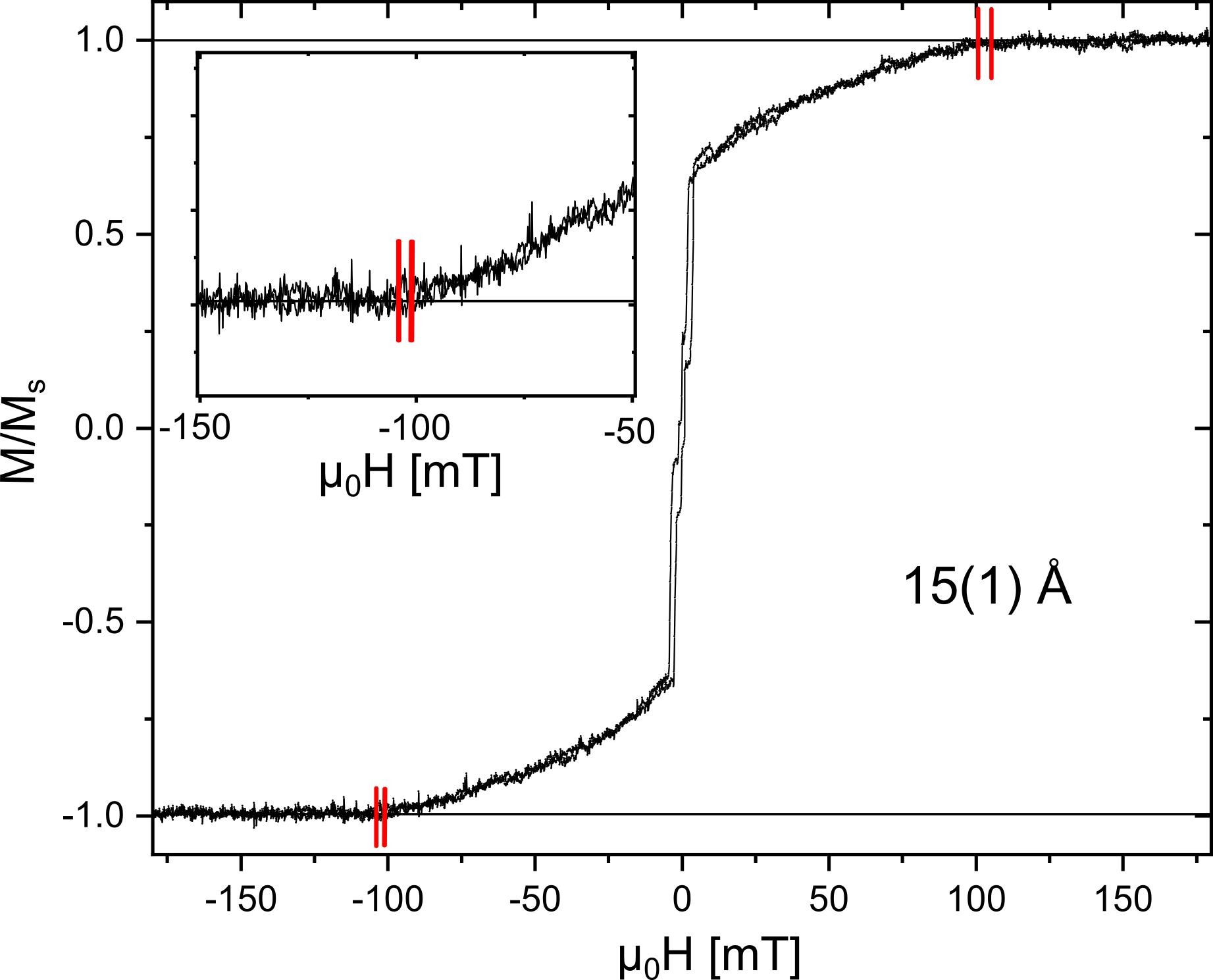}
\caption{Magnetic hysteresis loop of a [Fe/MgO]$_{10}$ superlattice measured with L-MOKE on a 10$\times$10~mm$^2$ sample with an Fe layer thickness 15(1)~{\AA}. The external magnetic field is applied along a Fe magnetic hard axis, i.e., Fe~[$110$]. The red lines indicate the error of the estimated saturation field.} 
\label{fig:sat_HA}\hfill
\end{figure*}

For each sample, hysteresis loops were measured with L-MOKE along the two hard axis directions, i.e., Fe~[$110$] and [$1\bar{1}0$]. The saturation field was determined from the intersection with a straight line at $M/M_S=1$. One example of this procedure is given in Fig.~\ref{fig:sat_HA}. 

For the determination of the saturation field along the easy axis, a similar procedure was followed, but the error was smaller as the switching of the magnetization direction from saturation is more abrupt. 

\providecommand{\noopsort}[1]{}\providecommand{\singleletter}[1]{#1}%
%

\end{document}